\documentstyle[12pt]{article}
\newcommand{\s}{\section}

\newcommand{\hs}{\hspace}
\newcommand{\ts}{\hs*}
\newcommand{\vs}{\vspace}
\newcommand{\e}{\enskip}
\newcommand{\q}{\quad}

\newcommand{\f}{\frac}
\newcommand{\dps}{\displaystyle}

\newcommand{\beqs}{\begin{equation}}
\newcommand{\feqs}{\end{equation}}
\newcommand{\beqa}{\begin{eqnarray}}
\newcommand{\feqa}{\end{eqnarray}}
\newcommand{\cC}{{\cal C}}
\newcommand{\cK}{{\cal K}}

\newcommand{\cP}{\hat{{\cal P}}}
\newcommand{\Pbf}{\hat{{\bf P}}}

\newcommand{\cH}{{\cal H}}
\newcommand{\cA}{{\cal A}}

\newcommand{\cs}{\bar{\phi}}

\newcommand{\pbrkt}[2]{[#1, #2]_{{\scriptsize PB}}}
\newcommand{\comm}[2]{[#1, #2]}

\newcommand{\braket}[2]{<#1\mid #2>}

\begin{document}

\ts{-18pt}{\Large \bf Uncertainty Relations and Quantum Effects of Constraints in Chern-Simons Theory}
\vs{18pt}\\
{\bf M. Nakamura}\footnote{E-mail:mnakamur@hamamatsu-u.ac.jp} \vs{4mm} \\
{\it Department of Business Adminitaration and Informatics, Hmamatsu University, Miyakoda-cho 1230, Kita-ku, Hamamatsu-shi, Shizuoka 431-2102, Japan} \vs{16mm}\\

\begin{abstract}
It is well known that Chern-Simons Theories are in the constrained systems and their total Hamiltonians become identically zero, because of their gauge invariance. While treating the constraints quantum mechanially, it will be expected taht there remain the quantum fluctuations due to the uncertainty principle. Using the projection operator method (POM) and the theory of dynamical constraints, such fluctuation terms are systematically derived in the case of Abelian Chern-Simons theory. It is shown that these terms produce the effective mass in the complex scalar fields coupled to the CS fields.  
\end{abstract}\vs{-9pt}
\ts{36pt}{\small PACS 03.70.+k - Theory of quantized fields}\\
\ts{36pt}{\small PACS 11.10.Ef - Lagrangian and Hamiltonian approach}\vs{6pt}\\
\s{Introduction}
\ts{12pt}Untill now, Chern-Simons (CS) theories have been enormously investigated in the 2+1 dimensional gauge theory and applied to the field theory like the quantum gravity and the certain planner condensed matter physics such as the anyon physics and the quantum Hall effect.\cite{Dunne} \cite{Deser} The CS theories are in the gauge theory and therefore are the constrained systems. Following the constrained Hamiltonian formalism\cite{TH} \cite{Faddeev}, then, the total Hamiltonians of the CS theories  become identically zero. Since the constraints in the CS theories are the bosonic second-class ones, however, there exist the uncertainty relations among the corresponding operators. Transferring to the quantum theory, therefore, the quantum fluctuations of these operators become enhanced\cite{NK}. Then, it will be expected that the quantum corrections due to the uncertainty principle remain in the total Hamiltonian. When the CS theories are coupled with the matter fields, further, these terms will be expected to give the corrections to the matter fields. In this paper, we investigate these conjectures in the case of the Abelian Chern-Simons theory and the complex scalar filed by using the projection operator method (POM)\cite{NM} and the ACCS-expansion formulas in the dynamical constraints\cite{NK}.\\
\ts{12pt}The present paper is organized as follows. In order to more clarify our process to quantize the gauge theory as the constrained system, in sect. 2, we review in detail the quantization of the pure Abelian CS theory, and the quantum correction terms due to the uncertainty principle are derived in the total Hamiltonian through the applying the ACCS-expansion formulas together with appropriate minimal unceratainty states. In sect. 3, the quantization of the complex scalar field coupled to the CS field is accomplished, and it is shown that the effective mass term appear in the total Hamiltonian of the matter field. The conclutions are given in sect. 4. 

\s{Quantization of Abelian Chern-Simons Theory}

\ts{12pt}In order to clarify the proccedere of the quantization of the Chern-Simons (CS) theory in terms of the POM, in this section, we attempt the quantization of the 2+1 dimensional Abelian Chern-Simons theory with $U(1)$ symmetry.\\
The Lagrangian density of the pure CS theory described by the gauge fields $A_{\mu}(x)\q (\mu=0,1,2)$ is given by 
$$
{\cal L}^0_{{\footnotesize CS}}=\f{\kappa}2\epsilon^{\mu\nu\rho}A_{\mu}\partial_{\nu}A_{\rho},
\eqno{(2.1)}
$$
where the metric tensor $g^{\mu\nu} =(+1,-1,-1)$ and $\epsilon^{012}=1$.

\subsection{Hamiltonian and constraints}

\ts{12pt}Let $L=\dps{\int}d^2x{\cal L}^0_{{\footnotesize CS}}$ be the Lagrangian of the system, then, the canonical momenta $\Pi^{\mu}(x)$ are defined by\footnote{the romanm indices $i,j,k,\cdots =1,2$ denote the spatial components.}
$$
\Pi^{\mu}=\f{\partial L}{\partial \dot{A}_{\mu}}=\f{\kappa}2\epsilon^{0\mu k}A_k.
\eqno{(2.2)}
$$
The canonical structure of the dynamical variables $A_{\mu}(x)$ and $\Pi^{\mu}(x)$ is given by 
$$
\begin{array}{rcl}
\pbrkt{A_{\mu}(x)}{A_{\nu}(y)}&=&\pbrkt{\Pi^{\mu}(x)}{\Pi^{\nu}(y)}=0,\vs{12pt}\\
\pbrkt{A_{\mu}(x)}{\Pi^{\nu}(y)}&=&g^{\nu}_{\mu}\delta^2(x-y),
\end{array}\eqno{(2.3)}
$$
where the Poisson bracket $\pbrkt{\q}{\q}$, and the delta function $\delta^2(x-y)=\delta(x^1-y^1)\delta(x^2-y^2)$.\\
Now, the canonical momenta (2.2) are the primary constraints 
$$
\chi_1^{\mu}=\Pi^{\mu}-\f{\kappa}2\epsilon^{0\mu k}A_k\approx 0.
\eqno{(2.4)}
$$
Then, $\chi_1^{\mu}(x)$ satisfy the Poisson bracket algebra
$$
\begin{array}{rcl}
\pbrkt{\chi_1^0(x)}{\chi_1^0(y)}&=&\pbrkt{\chi_1^0(x)}{\chi_1^k(y)}=0,\vs{12pt}\\
\pbrkt{\chi_1^k(x)}{\chi_1^l(y)}&=&-\kappa\varepsilon^{kl}\delta^2(x-y),
\end{array}
\eqno{(2.5)}
$$
where $\varepsilon^{12}=-\varepsilon^{21}=1$. So, it is convenient to classify the primary constraints into the following two:
$$
\begin{array}{rcl}
\chi_1^0&=&\Pi^0\approx 0,\vs{12pt}\\
\chi_1^k&=&\Pi^k-\dps{\f\kappa 2}\varepsilon^{kl}A_l\approx 0.
\end{array}
\eqno{(2.6)}
$$
Performing the Legendre transformation, then, the canonical Hamiltonian is given by
$$
H_0=\int dx^2(-\f\kappa 2\varepsilon^{kl}A_0\partial_kA_l+\f\kappa 2\varepsilon^{kl}\partial_kA_0A_l+u\chi_1^0+u_k\chi_1^k),\eqno{(2.7)}
$$
where the unknown Lagrange multipliers $u=\dot{A}_0$ and $u_k=\dot{A}_k$.\\
\ts{12pt}Now, the consistency conditions for the time evolusions of the primary cosntraints, $\dot{\chi}_1^{\mu}=\pbrkt{\chi_1^{\mu}}{H_0}\approx 0$, yield the secondary constraint
$$
\chi_2=-\dps{\f\kappa 2}\varepsilon^{kl}(\partial_kA_l-\partial_lA_k)\approx0
\eqno{(2.8)}
$$
and determine the Lagrange multipliers
$$
u_k=\partial_kA_0.
\eqno{(2.9)}
$$
\ts{12pt}Transferring to the quantum theory is accomplished through replacing the Poisson bracket into the commutator as follows:
$$
\pbrkt{\q }{\q }\e \longrightarrow\e \dps{\f{1}{i\hbar}}\comm{\q }{\q }.
\eqno{(2.10)}
$$
Let ${\cal C}^0_{\mbox{{\scriptsize P}}}=(A_{\mu}(x),\e \Pi^{\mu}(x))$ be the primary canonically conjugate set of the  field operators, which has the commutator algebra ${\cal A}^0_{\mbox{{\scriptsize P}}}({\cal C}^0_{\mbox{{\scriptsize P}}})$ given by
$$
\begin{array}{rcl}
\comm{A_{\mu}(x)}{A_{\nu}(y)}&=&\comm{\Pi^{\mu}(x)}{\Pi^{\nu}(y)}=0,\vs{12pt}\\
\comm{A_{\mu}(x)}{\Pi^{\nu}(y)}&=&i\hbar g^{\nu}_{\mu}\delta^2(x-y).
\end{array}
\eqno{(2.11)}
$$
The primary Hamiltonian becomes
$$
H_0=\int d^2x(-\f\kappa 2\varepsilon^{kl}(A_0\partial_kA_l+A_k\partial_lA_0)+\{u, \chi^0_1\}_{\mbox{\scriptsize S}}+\{u_k, \chi^k_1\}_{\mbox{\scriptsize S}})
\eqno{(2.12)}
$$
where the symmetrized product of any two operators $f$ and $g$
$$
\{f, g\}_{\mbox{\scriptsize S}}=\f12(fg+gf).
\eqno{(2.13)}
$$
The consistent set of constaints consists of 
$$
\begin{array}{rcl}
\chi_1^0&=&\Pi^0,\vs{12pt}\\
\chi_1^k&=&\Pi^k-\dps{\f\kappa 2}\varepsilon^{kl}A_l\vs{12pt}\\
\chi_2&=&-\dps{\f\kappa 2}\varepsilon^{kl}(\partial_kA_l-\partial_lA_k),
\end{array}
\eqno{(2.14)}
$$
which obeys the constraint algebra  
$$
\begin{array}{rcl}
\comm{\chi_1^0(x)}{\chi_1^0(y)}&=&\comm{\chi_1^0(x)}{\chi_1^k(y)}=\comm{\chi_1^0(x)}{\chi_2(y)}=0\vs{12pt}\\
\comm{\chi_1^k(x)}{\chi_1^l(y)}&=&-i\hbar\kappa\varepsilon^{kl}\delta^2(x-y),\vs{12pt}\\
\comm{\chi_2(x)}{\chi_2(y)}&=&0,\vs{12pt}\\
\comm{\chi_1^k(x)}{\chi_2(y)}&=&-i\hbar\kappa\varepsilon^{kl}\partial_l\delta^2(x-y).
\end{array}
\eqno{(2.15)}
$$
The constraint algebra (2.15) says that $\chi_1^0(x)$ is in the first class and $\chi_1^k$, $\chi_2$ are in the second class. 

\subsection{Sequential projection of operators} 

Following the POM, we perform the sequential transformations of the operators through introducing  a series of th projection operators.

\subsubsection{$\chi_1^k=0$ sector} 

From the constraint algebra (2.15), the {\it ACCS}\cite{NM} of the operators $\chi_1^k$ are given by
$$
\begin{array}{rcl}
\Theta(x)&=&\dps{\f1{\sqrt{2\kappa}}}(\chi_1^1(x)+\chi_1^2(x)),\vs{12pt}\\
\Xi(x)&=&\dps{\f1{\sqrt{2\kappa}}}(\chi_1^1(x)-\chi_1^2(x)),
\end{array}
\eqno{(2.16)}
$$
which satisfy the canonical commutaion relations (CCR)
$$
\begin{array}{rcl}
\comm{\Theta(x)}{\Theta(y)}&=&\comm{\Xi(x)}{\Xi(y)}=0,\vs{12pt}\\
\comm{\Theta(x)}{\Xi(y)}&=&i\hbar\delta^2(x-y).
\end{array}\eqno{(2.17)}
$$
Let $\cP^0_{\mbox{{\scriptsize I}}}$ be the projection operator with respect to the constraints $\chi_1^k=0$. Following the POM\cite{NM}, then, we  obtain the following results:\\
(i) Canonically conjugate set ${\cal C}^0_{\mbox{{\scriptsize I}}}$
$$
{\cal C}^0_{\mbox{{\scriptsize I}}}=\cP^0_{\mbox{{\scriptsize I}}}{\cal C}^0_{\mbox{{\scriptsize P}}}=(A_k(x), A_0(x),\Pi^0(x))\hs{12pt}\mbox{with}\q \Pi^k(x)=\dps{\f{\kappa}2}\varepsilon^{kl}A_l(x)
$$
(ii) Cmmutator\e algebra ${\cal A}^0_{\mbox{{\scriptsize I}}}({\cal C}^0_{\mbox{{\scriptsize I}}})$ 
$$
\begin{array}{rcl}
\comm{A_0(x)}{\pi^0(y)}&=&i\hbar\delta^2(x-y),\vs{6pt}\\
\comm{A_k(x)}{A_l(y)}&=&i\hbar\kappa^{-1}\varepsilon^{kl}\delta^2(x-y).
\end{array}
\eqno{(2.18)}
$$
(iii) The projected Hamiltonian $H_{\mbox{{\scriptsize I}}}$
$$
H_0\e \longmapsto H_{\mbox{{\scriptsize I}}}=\cP^0_{\mbox{{\scriptsize I}}}H_0=\int d^2x( A_0\chi_2 + \{u_0, \chi_1^0\}_{\mbox{\scriptsize S}}).
\eqno{(2.19)}
$$
(iv) The remaining constraints and cosntraint algebra
$$
\begin{array}{rcl}
\chi_1^0&=&0,\vs{12pt}\\
\chi_2&=&-\dps{\f\kappa 2}\varepsilon^{kl}(\partial_kA_l-\partial_lA_k)=0\hs{36pt}(\mbox{Gauss' law constraint})
\end{array}
\eqno{(2.20)}
$$ 
$$
\begin{array}{rcl}
\comm{\chi_1^0(x)}{\chi_1^0(y)}&=&\comm{\chi_2(x)}{\chi_2(y)}=\comm{\chi_1^0(x)}{\chi_2(y)}=0\vs{12pt}\\
\comm{\chi_1^0(x)}{H_{\mbox{{\scriptsize I}}}}&=&\comm{\chi_2(x)}{H_{\mbox{{\scriptsize I}}}}=0.
\end{array}
\eqno{(2.21)}
$$
Then, the remaining operator-constraints are in the first class, and the Hamiltonian is invariant under the the gauge transformation
$$
A_k\e \longmapsto \e A_k'=A_k-\partial_k g.
\eqno{(2.22)}
$$
The projection of field operators with respect to the first class constraints are performed with two ways, one of which is the gauge-fixing way, and the other, the gauge-unfixing one.

\subsubsection{$\chi_1^0 = 0$ sector}

(1) {\bf The gauge fixing case}\\
\ts{12pt} As the gauge fixing condition, we adopt the Weyl gauge
$$
A_0(x)=0.
\eqno{(2.23)}
$$
Let $\cP^0_{\mbox{{\scriptsize II}}}$ be the projection operator for the {\it ACCS} $(A_0(x), \chi_1^0)$. Then, we obtain the following results.
\begin{description}
\item{(i)}\hs{12pt}${\cal C}^0_{\mbox{{\scriptsize II}}}=\cP^0_{\mbox{{\scriptsize II}}}{\cal C}^0_{\mbox{{\scriptsize I}}}=(A_k(x))$
\item{(ii)}\hs{12pt}${\cal A}^0_{\mbox{{\scriptsize II}}}({\cal C}^0_{\mbox{{\scriptsize II}}})={\cal A}^0_{\mbox{{\scriptsize I}}}({\cal C}^0_{\mbox{{\scriptsize II}}})$
\item{(iii)}\hs{12pt}$H_{\mbox{{\scriptsize II}}}=\cP^0_{\mbox{{\scriptsize II}}}H_{\mbox{{\scriptsize I}}}=0$.\\
\end{description}
(2) {\bf The gauge unfixing case}\\
\ts{12pt} Let the {\it ACCS} be $(A_0(x), \chi_1^0(x))$. Then, the projection operator, $\cP^0_{{\scriptsize \chi}}$, eliminating the constraint-operator $\chi_0^1(x)$ is given by
$$
\cP^0_{{\scriptsize \chi}}=\sum^{\infty}_{n=0}\f{(-)^n}{n!}(\hat{\chi}_0^{1(+)})^n(\hat{A}_0^{(-)})^n,
\eqno{(2.24)}
$$
which satisfies
$$
\cP^0_{{\scriptsize \chi}}\chi_0^1=0,\hs{36pt}\cP^0_{{\scriptsize \chi}}A_0=A_0.
\eqno{(2.25)}
$$
Then, the following results are given.
\begin{description}
\item{(i)}\hs{12pt}${\cal C}^0_{\mbox{{\scriptsize II'}}}=\cP^0_{{\scriptsize \chi}}{\cal C}^0_{\mbox{{\scriptsize I}}}=(A_k(x), A_0(x))$
\item{(ii)}\hs{12pt}${\cal A}^0_{\mbox{{\scriptsize II'}}}({\cal C}^0_{\mbox{{\scriptsize II'}}})={\cal A}^0_{\mbox{{\scriptsize I}}}({\cal C}^0_{\mbox{{\scriptsize II'}}})$
\item{(iii)}\hs{12pt}$H_{\mbox{{\scriptsize II'}}}=\cP^0_{{\scriptsize \chi}}H_{\mbox{I}}=\dps{\int} d^2x A_0(x)\chi_2(x)$, 
\end{description}
where $A_0(x)$ is the c-number field.

\subsubsection{$\chi_2=0$ sector}

For the Gauss' law constrtaint $\chi_2=-\dps{\f\kappa 2}\varepsilon^{kl}(\partial_kA_l-\partial_lA_k)$, the {\it ACCS} is given as
$$
\begin{array}{rcl}
\xi(x)&=&n^kA_k(x),\vs{12pt}\\
\varpi(x)&=&-\dps{\int}d^2zg^A(x,y)\chi_2(y),
\end{array}
\eqno{(2.26)}
$$
where $n_k$ is the c-number operator with $n_kn^k=1$, and the 2-point function $g^A(x,y)$  is defined as follows:
$$
g^A(x,y)=n_1\epsilon(x^1,y^1)\delta(x^2-y^2)+n_2\delta(x^1-y^1)\epsilon(x^2,y^2)
\eqno{(2.27)}
$$
with $\epsilon(x,y)=\dps{\f12}(\theta(x-y)-\theta(y-x))$ ($\theta(x)$ : the step function).
Then, $\xi(x)$ and $\varpi(x)$ obey the CCR
$$
\begin{array}{l}
\comm{\xi(x)}{\xi(y)}=\comm{\varpi(x)}{\varpi(y)}=0\vs{6pt}\\
\comm{\xi(x)}{\varpi(y)}=i\hbar\delta^2(x-y).
\end{array}
\eqno{(2.28)}
$$
\\      
{\bf (1) gauge fixing case}\\
\ts{12pt} The gauge condition $n^kA_k(x)=0$ is the axial gauge in the $n^k$ direction. Let $\cP^{A}$ be the projection operator of the {\it ACCS} $\xi$ and $\varpi$.  Then, we obtain the following results.
\begin{description}
\item{(i)}\hs{12pt}${\cal C}^0_{\mbox{{\scriptsize A}}}=\cP^A{\cal C}^0_{\mbox{{\scriptsize II'}}}=(A_k(x), A_0(x))$\\
 with $\varepsilon^{kl}\partial_kA_l(x)=0$ and the c-number field $A_0(x)$,
\item{(ii)}\hs{12pt}${\cal A}_A({\cal C}^0_{\mbox{{\scriptsize A}}})$ :\\
$$
\comm{A_k(x)}{A_l(y)}=0,
\eqno{(2.29)}
$$
\item{(iii)}\hs{12pt}$H_{\mbox{{\scriptsize A}}}=\cP^AH_{\mbox{{\scriptsize II'}}}=0$. \\
\end{description}
{\bf (2) gauge unfixing case}\\
The projection operator for the Gauss'law constraint $\chi_2=0$ is given by 
$$
\cP_{\varpi}=\sum^{\infty}_{n=0}\f{(-1)^n}{n!}(\hat{\varpi}^{(+)})^n(\hat{\xi}^{(-)})^n.
\eqno{(2.30)}
$$
Then, we obtain the equivalent results to the gauge fixing case, besides $n^kA_k=c$.

\subsection{{\it ACCS}-expansion of CS theory}

Let the {\it ACCS} be $(Q(x), P(x))$. Then, the {\it ACCS}-expansion of any field operator $O(x)$ is written as\cite{NK}
$$
O(x)=\sum^{\infty}_{n,m=0}\f{(-1)^n}{n!m!}(\hat{Q}^{(+)})^n(\hat{P}^{(+)})^m\cP(\hat{Q}^{(-)})^m (\hat{P}^{(-)})^n O(x),
\eqno{(2.31)}
$$
where 
$$
(\hat{F}^{(+)})^n\cdots(\hat{G}^{(-)})^n=\int d^2z_1\cdots d^2z_n\hat{F}^{(+)}(z_1)\cdots\hat{F}^{(+)}(z_n)\cdots \hat{G}^{(-)}(z_n)\cdots \hat{G}^{(-)}(z_1).
$$
Let $\cH^c$ be the subspace of the Hilbert space $\cH$, on which the {\it ACCS} operates\cite{NK}, and  we introduce the {\it hyper}-operator $\Pbf$ as follows: 
$$
\Pbf O(x) = \f{<\Phi\mid O(x)\mid\Phi>}{<\Phi\mid\Phi>}
\eqno{(2.32)}
$$
with 
$$
\Phi \in \cH^c.
\eqno{(2.33)}
$$
Then, the effective Hamiltonian $H^{\mbox{eff}}$, which contain the quantum corrections due to the uncertainty principle, is given by 
$$
H^{\mbox{eff}}=\Pbf H=\cP H + H^{\mbox{eff}}_{\mbox{qc}},
\eqno{(2.34)}
$$
where the additional term is defined by
$$
H^{\mbox{eff}}_{\mbox{qc}}=\sum^{\infty}_{n+m\neq 0} \f{(-1)^n}{n!m!}<(\hat{Q}^{(+)})^n(\hat{P}^{(+)})^m\cdot 1>_{\Phi}\cP(\hat{Q}^{(-)})^m (\hat{P}^{(-)})^n H,
\eqno{(2.35)}
$$
where
$$
<(\hat{Q}^{(+)})^n(\hat{P}^{(+)})^m\cdot 1>_{\Phi}=\Pbf (\hat{Q}^{(+)})^n(\hat{P}^{(+)})^m\cdot 1.
\eqno{(2.36)}
$$
The ground state in a minimal uncertainty state is appropriate to be adopted as the state vector $\Phi \in \cH^c$.\cite{NK}. Becaude of the linearity of the field operators, then, the additional term $H^{\mbox{eff}}_{\mbox{qc}}$ in the pure Abelian CS-theory becomes zero. So, the effective total Hamiltonian also becomes zero.

\section{Quantum effect in complex scalar fields}

\subsection{Quantization of complex scalar field in the Abelian CS-theory}

The Lagrangian of the complex scalar field $\phi(x),\phi^*(x)$ coupled to the Abelian CS fields is given by 
$$
L=\int d^2x(\overline{D_{\mu}\phi}D^{\mu}\phi + \f{\kappa}2\epsilon^{\mu\nu\rho}A_{\mu}\partial_{\nu}A_{\rho}),
\eqno{(3.1)}
$$
which is invariant under the U(1)-gauge transformation with $U=e^{ig(x)}$, where the covariant derivative $D_{\mu}=\partial_{\mu}-iA_{\mu}(x)$ and $\cs=\phi^*$. Following the program shown in the previous section, we perform the quantization of the system.\\
The canonical momenta are defined as follows:
$$
\begin{array}{rcl}
\pi(x)&=&\dps{\f{\partial L}{\partial \dot{\phi}(x)}}=\dot{\phi}^*(x)+iA_0(x)\phi^*(x)\vs{12pt}\\
\pi^*(x)&=&\dps{\f{\partial L}{\partial \dot{\phi}^*(x)}}=\dot{\phi}(x)-iA_0(x)\phi^*(x)\vs{12pt}\\ 
\Pi^{\mu}&=&\dps{\f{\partial L}{\partial \dot{A}_{\mu}}=\f{\kappa}2\epsilon^{0\mu k}A_k}.
\end{array}
\eqno{(3.2)}
$$
Then, we obatin the following the constrained Hamiltonian system $(\cC_{\mbox{\scriptsize P}} ; H_{\mbox{\scriptsize P}} ; \cK_{\mbox{\scriptsize P}})$:
$$
\begin{array}{rcl}
\cC_{\mbox{\scriptsize P}}&=&\{(\phi,\pi),(\phi^*,\pi^*),(A_{\mu},\Pi^{\mu})\},\vs{12pt}\\
H_{\mbox{\scriptsize P}}&=&\dps{\int d^2x(\pi^*\pi-\overline{(D\phi)}\cdot(D\phi)+A_0\chi_G+u_{\mu}\chi^{\mu})},\vs{12pt}\\
\cK_{\mbox{\scriptsize P}}&=&\{\chi^{\mu}, \chi_G\},
\end{array}
\eqno{(3.3)}
$$
where $\overline{(D\phi)}\cdot(D\phi)=\overline{(D^k\phi)}(D_k\phi)$ and the consistent set of constraints $\cK_{\mbox{\scriptsize P}}$ :
$$
\begin{array}{rcl}
\chi^{\mu}&=&\Pi^{\mu}-\dps{\f{\kappa}2}\epsilon^{0\mu\nu}A_{\nu}\approx 0,\vs{6pt}\\
\chi_G&=&j_0-\kappa\epsilon^{kl}\partial_kA_l\approx 0\q \mbox{with}\e j_0=i(\pi\phi-\bar{\phi\pi})\hs{12pt}\mbox{(Gauss' law constraint)}.
\end{array}
\eqno{(3.4)}
$$
The primary commutator algebra $\cA_{\mbox{\scriptsize P}}(\cC_{\mbox{\scriptsize P}})$ is represented as
$$
\begin{array}{l}
\comm{\phi(x)}{\pi(y)}=\comm{\phi^*(x)}{\pi^*(y)}=i\hbar\delta^2(x-y),\vs{6pt}\\
\comm{A_{\mu}(x)}{\Pi^{\nu}(y)}=i\hbar g^{\nu}_{\mu}\delta^2(x-y),\vs{6pt}\\
\mbox{(the others)}=0.
\end{array}
\eqno{(3.5)}
$$
Eliminating the second-class constraint-operators $\chi^k=\Pi^k-\dps{\f{\kappa}2}\epsilon^{kl}A_l$ and introducing the Weyl gauge $A_0(x)=0$, then, $(\cC_{\mbox{\scriptsize P}} ; H_{\mbox{\scriptsize P}} ; \cK_{\mbox{\scriptsize P}})$ is transformed to the constrained system $(\cC_{\mbox{\scriptsize S}} ; H_{\mbox{\scriptsize S}} ; \cK_{\mbox{\scriptsize S}})$ as follows:
$$
\begin{array}{rcl} 
\cC_{\mbox{\scriptsize P}}&\mapsto& \cC_{\mbox{\scriptsize S}}=\{(\phi,\pi),(\phi^*,\pi^*),(A_k)\},\vs{12pt}\\
H_{\mbox{\scriptsize P}}&\mapsto&H_{\mbox{\scriptsize S}}=\dps{\int d^2x(\pi^*\pi-\overline{(D\phi)}\cdot(D\phi))},\vs{12pt}\\
\cK_{\mbox{\scriptsize P}}&\mapsto&\cK_{\mbox{\scriptsize S}}=\{\chi_{\mbox{\scriptsize G}}\}.
\end{array}
\eqno{(3.6)}
$$ 
The commutator-algebra of $\cC_{\mbox{\scriptsize S}}$ is given by
$$
\begin{array}{l}
\comm{\phi(x)}{\pi(y)}=\comm{\phi^*(x)}{\pi^*(y)}=i\hbar\delta^2(x-y),\vs{6pt}\\
\comm{A_k(x)}{A_l(y)}=i\hbar\kappa^{-1}\epsilon^{kl}\delta^2(x-y),\vs{6pt}\\
\mbox{(the others)}=0.
\end{array}
\eqno{(3.7)}
$$
As well as in the pure Abelian CS theory, finally, the constrained system satisfying the Gauss' law $\kappa^{kl}\partial_kA_l(x)=j_0(x)$ becomes as follows:
$$
(\cC_{\mbox{\scriptsize S}} ; H_{\mbox{\scriptsize S}} ; \cK_{\mbox{\scriptsize S}})\mapsto(\cC_{\mbox{\scriptsize G}} ; H_{\mbox{\scriptsize G}}),
\eqno{(3.8)}
$$
where
$$
\begin{array}{rcl}
\cC_{\mbox{\scriptsize G}}&=&\{(\phi,\pi),(\phi^*,\pi^*),(A_k)\}\vs{6pt}\\
                          & &\mbox{with}\hs{12pt} n^kA_k(x)=0,\hs{12pt}\kappa\varepsilon^{kl}\partial_kA_l(x)=j_0(x),\vs{12pt}\\
H_{\mbox{\scriptsize G}}&=&\dps{\int d^2x(\pi^*\pi-\overline{(D\phi)}\cdot(D\phi))}.
\end{array}
\eqno{(3.9)}
$$
Under the axial gauge, the commutator of $A_k(x)$ becomes zero,
$$
\comm{A_k(x)}{A_l(y)}=0.
\eqno{(3.10)}
$$

\subsection{Quantum Fluctuations of Constraints} 

When introducing the ground state of the minimal uncertainty states  as the state vector $\Phi \in \cH^c$ in the additional term (2.36), the quantum correction term is produced from $\dps{\int d^2x A^k(x)A_k(x)\phi^*(x)\phi(x)}$. In the case of the coherent state,
$$
\braket{Q}{\Phi} = \left(\f{1}{\pi \hbar}\right)^{1/4}\exp[-\f1{2\hbar}Q\cdot Q],
\eqno{(3.11)}
$$
the additional term $H^{\mbox{eff}}_{\mbox{qc}}$ in the constrained system $(\cC_{\mbox{\scriptsize S}} ; H_{\mbox{\scriptsize S}} ; \cK_{\mbox{\scriptsize S}})$ becomes
$$
H^{\mbox{eff}}_{\mbox{qc}}=\int d^2x\f{\hbar}{2\kappa}\phi^*\phi,
\eqno{(3.12)}
$$
which produces the effective mass of the scalar field. Taking account of the {\it Gauss' law} sector, further, it becomes 
$$
H^{\mbox{eff}}_{\mbox{qc}}=\int d^2x{\hbar}(1+\f1{2\kappa})\phi^*\phi .\eqno{(3.13)}
$$
Thus, we obatain the effective Hamiltonian containing the quantum corrections caused by the uncertainty relations of the constraint-operators
$$
H^{\mbox{eff}}=\int d^2x(\pi^*\pi-\overline{(D\phi)}\cdot(D\phi)+\mu^2\phi^*\phi),
\eqno{(3.14)}
$$
where
$$
\mu^2=\hbar(1+\f1{2\kappa}).
\eqno{(3.15)}
$$

\s{Conclusions}

We have investigated the quantum corrections of constraints in gauge theories when imposing the constraints in the operator form. In order to forcus the quantum corrections due to the uncertainty relations among the constraint-operators, we have taken the pure Abelian CS theory and the complex scalar fields coupled to the CS theory as the gauge-invariant system. Then, we have obtained the following results.\\

\ts{-12pt}(i) Because of the linearity in the gauge-fields, the pure Abelian CS theory has no correction due to the uncertainty principle.\\
(ii)  In the complex scalar field system, the quantum fluctuations of the gauge fields produce the effective mass in the matterl field.\\
 
Then, it  will be expected that it is one of the most interesting problems in hte gauge theory to investigate the relation of our mechanism to produce the effective mass with the Higgs mechanism.\cite{Higgs}

\end{document}